\documentclass[conference]{IEEEtran}
\IEEEoverridecommandlockouts
\usepackage{cite}
\usepackage{amsmath,amssymb,amsfonts}
\usepackage{algorithmic}
\usepackage{graphicx}
\usepackage{textcomp}
\usepackage{xcolor}
\usepackage{booktabs}
\usepackage{underscore}
\usepackage{tabularx}
\usepackage{array}
\usepackage{authblk}  
\usepackage{hyperref}
\newcolumntype{L}{>{\raggedright\arraybackslash}X}

\def\BibTeX{{\rm B\kern-.05em{\sc i\kern-.025em b}\kern-.08em
    T\kern-.1667em\lower.7ex\hbox{E}\kern-.125emX}}
\begin{document}

\title{An Autonomous Agent Framework for Feature-Label Extraction from Device Dialogues and Automatic Multi-Dimensional Device Hosting Planning Based on Large Language Models}


\author{%
	Huichao Men$^{1}$, Yizhen Hu$^{1,2}$, Yu Gao$^{1,\ast}$, Xiaofeng Mou$^{1}$, Yi Xu$^{1,\ast}$, Xinhua Xiao$^{1}$\\
	\small $^{1}$Midea AI Research Center, Shanghai, P.R. China\\
	\small $^{2}$Harbin Engineering University, Harbin, P.R. China\\
	\small $^{\ast}$Corresponding authors\\
	\small \href{mailto:menhuichao@midea.com}{menhuichao@midea.com}, 
	\small \href{mailto:huyz64@midea.com}{huyz64@midea.com}, 
	\small \href{mailto:gaoyu11@midea.com}{gaoyu11@midea.com}, 
	\small \href{mailto:mouxf@midea.com}{mouxf@midea.com}, 
	\small \href{xuyi42@midea.com}{xuyi42@midea.com},
	\small \href{xiaoxh3@midea.com}{xiaoxh3@midea.com},
}

\maketitle

\begin{abstract}
With the deep integration of artificial intelligence and smart home technologies, the intelligent transformation of traditional household appliances has become an inevitable trend. This paper presents AirAgent—an LLM-driven autonomous agent framework designed for home air systems. Leveraging a voice-based dialogue interface, AirAgent autonomously and personally manages indoor air quality through comprehensive perception, reasoning, and control. The framework innovatively adopts a two-layer cooperative architecture: “Memory-Based Tag Extraction” and “Reasoning-Driven Planning.” First, a dynamic memory tag extraction module continuously updates personalized user profiles. Second, a reasoning-planning model integrates real-time environmental sensor data, user states, and domain-specific prior knowledge (e.g., public health guidelines) to generate context-aware decisions. To support both interpretability and execution, we design a semi-streaming output mechanism that uses special tokens to segment the model’s output stream in real time, simultaneously producing human-readable Chain-of-Thought (CoT) explanations and structured, device-executable control commands. The system handles planning across 25 distinct complex dimensions while satisfying over 20 customized constraints. As a result, AirAgent endows home air systems with proactive perception, service, and orchestration capabilities, enabling seamless, precise, and personalized air management responsive to dynamic indoor and outdoor conditions. Experimental results demonstrate up to 92.5\% accuracy and a >20\% improvement in user experience metrics compared to competing commercial solutions.
\end{abstract}


\section{Memory Tag Extraction Module}
\subsection{Data Construction}
To train the model with fine-grained semantic understanding and entity extraction capabilities, we construct an instruction-tuning dataset in the “Instruction–Input–Output” format. The core of this data construction lies in parsing complex pronoun references and nuanced expressions related to users’ physical states and thermal preferences.

To ensure consistency and standardization in tag extraction, this work establishes a hierarchical tagging schema covering four dimensions: user demographics, environmental preferences, health conditions, and memory lifecycle management:

\subsubsection{Population Group (populationGroup)}
Household members are categorized into “Adult Male,” “Adult Female,” “Child,” “Elderly,” and “Other,” enabling the system to align basic comfort models with physiological characteristics specific to age and gender.
\subsubsection{Thermal Preference (hotcoldPreference)}
User sensitivity to ambient temperature is represented on a five-point scale: “Very Cold-Sensitive,” “Slightly Cold-Sensitive,” “Neutral,” “Slightly Heat-Sensitive,” and “Very Heat-Sensitive.”
\subsubsection{Health Condition (physicalCondition)}
This dimension focuses on physiological states highly relevant to air quality. Predefined tags include “Common Cold,” “Fever,” “Cough,” “Rhinitis,” “Asthma,” and “Menstruation.”
\subsubsection{Memory State Management (action)}
To address the temporal validity of user states, a “memory expiration” trigger mechanism is introduced. When utterances conveying recovery—such as “recovered,” “feeling better,” or “all good now”—are detected, the model marks the corresponding health condition as expired, thereby removing outdated illness records from the user profile and preventing the continued execution of obsolete strategies.

\begin{table*}[ht]
	\centering
	\caption{Memory Tag Structure}
	\label{tab:tag_schema}
	\begin{tabular}{ll}
		\toprule
		\textbf{Category} & \textbf{Tag Values} \\
		\midrule
		Population Group     & Adult Male \\
		& Adult Female \\
		& Children \\
		& Elderly \\
		& Others \\
		\addlinespace
		Hot-Cold Preference  & Slightly Cold-Sensitive \\
		& Very Cold-Sensitive \\
		& Slightly Heat-Sensitive \\
		& Very Heat-Sensitive \\
		& Neutral \\
		\addlinespace
		Physical Condition   & Cold \\
		& Fever \\
		& Cough \\
		& Rhinitis \\
		& Asthma \\
		& Menstruation \\
		\addlinespace
		Recovery Status Management & Recovered, Healed, Fully Recovered, Back to Normal, All Clear \\
		\bottomrule
	\end{tabular}
\end{table*}

\begin{figure}[htbp]
	\centering
	\includegraphics[width=0.8\linewidth]{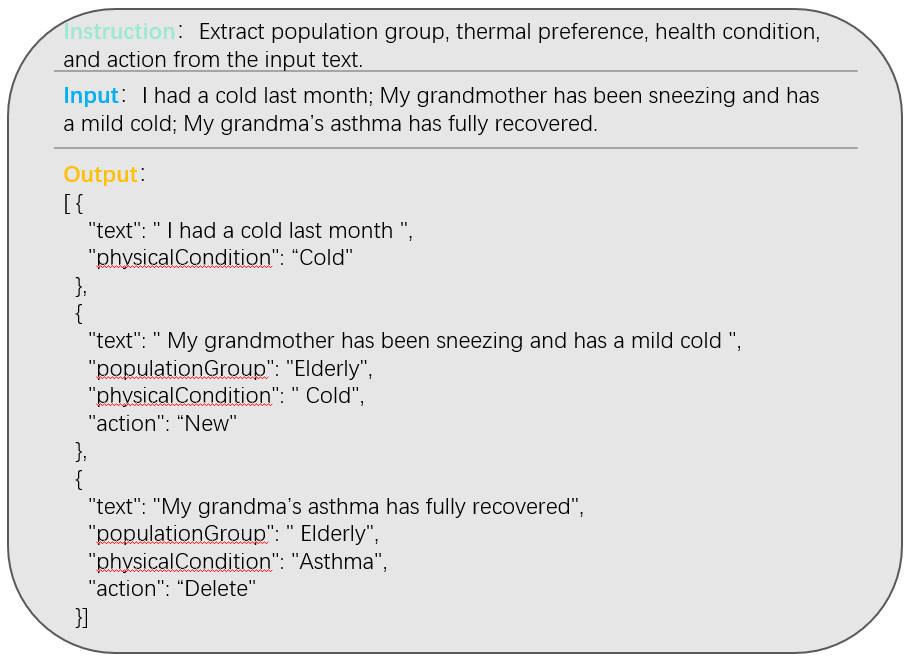}
	\caption{Data Construction for Memory Tag Extraction Model}
	\label{fig:method_overview}
\end{figure}

\subsection{Model Structure}
AirAgent’s memory-tag extraction pipeline first transcribes the spoken command into text via ASR, then feeds the text to an instruction-tuned memory model that performs entity extraction and state judgment, finally converting the results into structured user-memory profiles. The workflow consists of three core stages: voice interaction, semantic parsing, and user-profile maintenance; the detailed processing logic is as follows.

\subsubsection{Speech Interaction and Text Serialization}
The pipeline begins with the Automatic Speech Recognition (ASR) module listening to spontaneous user commands or daily conversations in the home environment. The ASR component converts the continuous audio stream into a text sequence and screens for utterances that pertain to the user’s health status, supplying the downstream stages with their initial input.

\subsubsection{Instruction-SFT Semantic Parsing}
After ASR, the text is forwarded to a memory-tag extraction model that has been instruction-SFT on a task-specific dataset. Driven by engineered prompts, the model performs deep semantic analysis and entity extraction rather than rudimentary keyword spotting, leveraging context to capture inter-entity dependencies. For instance, given the input “Grandma’s asthma has cleared up,” it not only identifies the entity “elderly” and the condition “asthma,” but—crucially—detects the state-change cue “has cleared up,” producing a parse annotated with "action": "delete/_condition". The model ultimately emits a standardized list of JSON objects that map the unstructured text to the predefined variables populationGroup, physicalCondition, hotColdPreference, etc.

\subsubsection{Dynamic Construction and Maintenance of User Profiles}
Guided by the action field in the model’s output, the system performs corresponding operations on the profile database. When "action": "add_condition" is detected, the relevant health tag is written into the current health-status memory of the corresponding family member; when "action": "remove_condition" is found, the obsolete status is purged. Through this loop AirAgent keeps a high-fidelity, real-time panorama of every household member’s state, furnishing downstream reasoning models with accurate input for personalized air-quality control decisions.

\section{Inference and Planning Module}
The inference-and-planning model serves as AirAgent’s core decision-making brain; its data construction focuses not only on fusing multi-source information, but—more critically—on keeping the planning and delegation process explainable and logically closed-loop.

\subsection{Data Construction}
Dataset construction for the inference model combines “data assembly” with a “semi-automatic chain-of-thought” pipeline, ensuring the model emits both user-optimized reasoning chains and logically valid control commands.

The input layer of the inference model integrates four data dimensions to create a panoramic perception of the current home environment.

\subsubsection{Objective Environment}
Fused spatio-temporal data (city, time-of-day) and meteorological data: outdoor weather (temperature, humidity, PM2.5) plus real-time indoor readings from sensors (temperature, humidity, CO2, TVOC, PM2.5, HCHO, etc.).
\subsubsection{Personalized User Profile}
Long- and short-term memories retrieved by the memory model: thermal preference (e.g., “prefers cool”), demographic group, and health conditions (e.g., “asthma”).
\subsubsection{Device Status}
Physical location (“bedroom”), current operating state (power, mode, set-point, fan speed, airflow feel), and available add-ons (“sterilize”, “purify”, etc.).
\subsubsection{Disease-Control Knowledge}
Expert knowledge base for the current season or prevalent respiratory illnesses.

\subsection{Inference-Model Tag System}
To give the inference model a panoramic awareness of the environment and precise execution boundaries, we design a two-dimensional tagging system that spans the device side and the environment side, mapping physical-world information into structured semantics the model can understand:

\subsubsection{Device Descriptor}
It defines the agent’s physical capability limits and current operating baseline, guaranteeing that every decision is executable.  
(1) Device Location: physical space label (e.g., “master bedroom”, “living room”).  
(2) Add-on Functions: feature labels such as “sterilize”, “purify”, “fresh-air”.  
(3) Device State: real-time snapshot labels for power on/off, operating mode, fan speed, etc.; these serve as the initial state for subsequent state-transition reasoning.
\subsubsection{Objective Environment}
The external input variables for decision-making, fusing spatio-temporal context with multi-dimensional meteorological data.  
(1) City \& Time-stamp: geo and temporal anchors that let the model associate region-specific climatic traits.  
(2) Outdoor Weather: live meteorological service data—outdoor temperature, humidity, PM2.5 concentration, etc.  
(3) Indoor Air: sensor-array readings—temperature, humidity, CO2, TVOC, PM2.5—directly trigger dynamic adjustment logic.

\subsection{Output of the Model}
To address the lack of interpret ability in traditional black-box models, the output structure designed in this paper employs a “semi-streaming output” mechanism. As shown in Figure 4, <special tokens> are introduced to divide the output into two parts: an explicit reasoning process and executable instructions. In the structured control-instruction segment of the proposed algorithm, the output must plan and adjust 25 attribute dimensions and automatically regulate the duration of different functions.

\subsubsection{Explicit Reasoning Chain}
This section presents the model’s thought process to enhance user experience. Data annotation follows the chain below:
(1) Environment \& User-State Perception  
Analyze current indoor–outdoor temperature difference, airborne pollutants (e.g., excess formaldehyde), and user health traits (e.g., asthma).
(2) Goal Setting  
Establish the priority of multi-objective optimization—for instance, satisfying “low airflow for asthma” while addressing “severe formaldehyde excess.”
(3) Quantitative-Target Determination  
Deduce ideal thresholds for each metric (e.g., CO2 < 800 ppm, PM2.5 < 15 µg/m³).
(4) Strategy Formulation  
Generate a concrete function-combination plan (e.g., “heating mode + purification on”).
(5) Reasoning \& Scheduling  
For high-power or special functions (e.g., sterilization), plan an intermittent runtime schedule.

\subsubsection{Structured Control Instructions}
This section is machine-readable JSON that directly drives device execution and contains:
(1)cmd: immediate control parameters for the device (mode, temperature, fan speed, airflow feel, etc.)  
(2)threshold: target environmental-quality thresholds used as the reference baseline for closed-loop control  
(3)interval_time: task-scheduling cycle for auxiliary functions (e.g., “sterilize for 30 min every 4 h”)  
(4)tips: natural-language explanations generated for the user, summarizing the decision rationale and current environmental risks

\begin{figure}[htbp]
	\centering
	\includegraphics[width=0.8\linewidth]{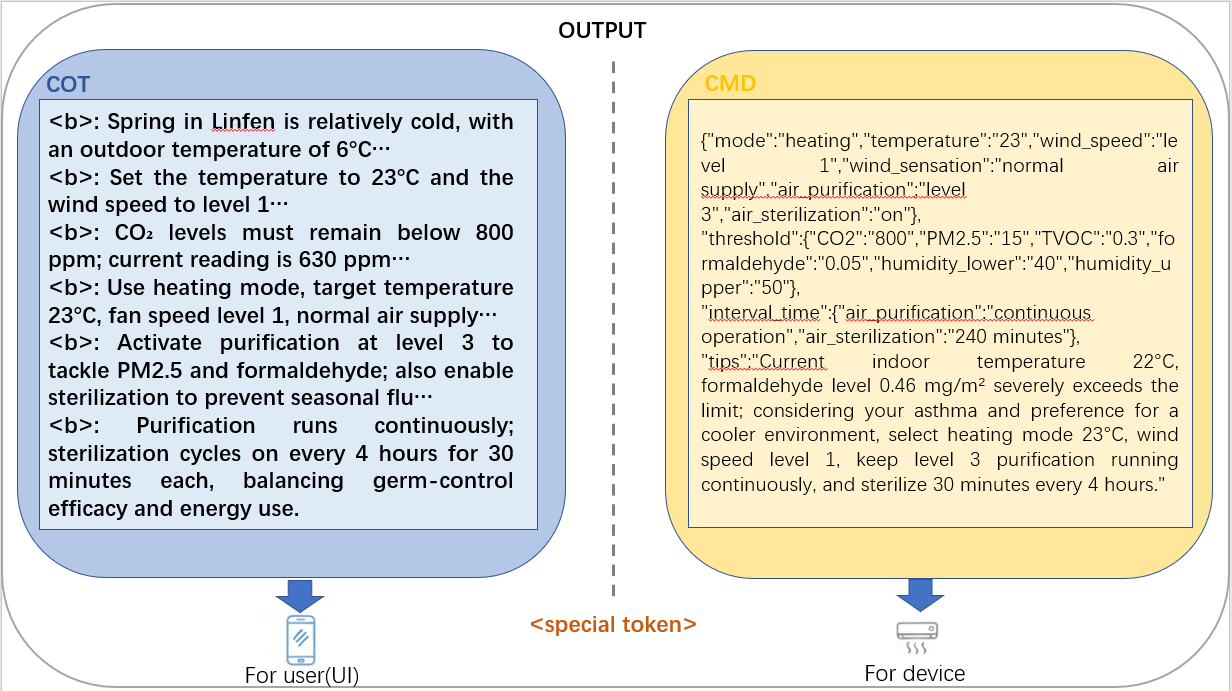}
	\caption{Reasoning-planning model semi-streaming output}
	\label{fig:reasoning_overview}
\end{figure}

\subsection{Model Structure}
As illustrated in Figure 5, the AirAgent architecture proposed in this paper follows a closed-loop control paradigm of “memory-tag extraction → reasoning \& planning,” aiming to leverage the semantic understanding and reasoning power of a large language model (LLM) together with relevant algorithmic policies to tackle the dual pain-points of home air systems: high cognitive load on users (“don’t know how to use”) and sluggish dynamic response (“too late to adjust”). Embedded in the air-system devices, AirAgent achieves full-dimensional perception and autonomous governance of up to 25 indoor parameters—temperature, humidity, airflow, cleanliness, freshness, etc. By coordinating a foundation-model-based fine-tuned module with algorithmic-policy modules, and by fusing multi-source inputs while decoupling outputs in a semi-streaming fashion, it forms a highly self-adaptive, proactive “air-steward brain” closed loop.
\begin{figure}[htbp]
	\centering
	\includegraphics[width=0.8\linewidth]{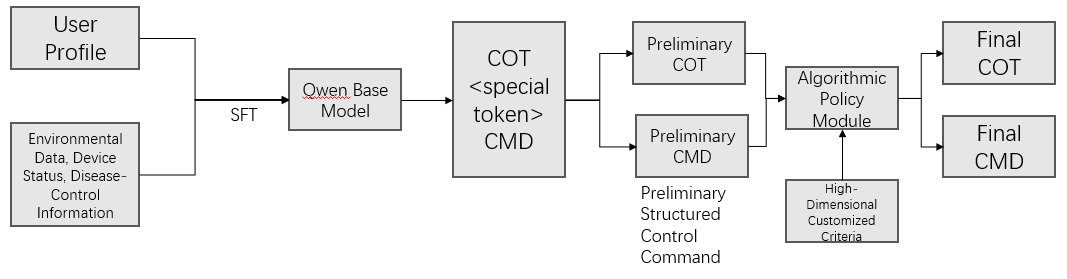}
	\caption{Method of Inference and Planning Module}
	\label{fig:planning_overview}
\end{figure}

\section{Experiments}
The experiments in this paper consist of two parts, each validating a respective module. Both are conducted on a complex dialogue dataset related to air systems, together with intricate environmental and device-status data.  
The current experimental status is presented module by module below.

\subsection{Experiments of Memory Tag Extraction Module}
This section presents experiments based on a competitor’s closed-source model and our in-house developed model. The primary evaluation criteria are: (1) a user experience (UX)-oriented standard defined by professional testers, which assesses whether each attribute in the output conforms to the corresponding attribute definition in the test cases; and (2) per-inference latency.
Each test sample contains an average of three user queries (actual counts may be higher or lower); all reported figures are therefore averaged over roughly three successive inferences of the module.

\begin{table*}[htbp]
	\centering
	\caption{Performance Comparison Between Competitor's Model and In-house Model}
	\label{tab:performance_comparison}
	\begin{tabular}{lcc}
		\toprule
		\textbf{Metric} & \textbf{Competitor's Model} & \textbf{In-house Model} \\
		\midrule
		User Experience Pass Rate & 60\% & 83.3\% \\
		Average Inference Latency & 0.51 s & 3.33 s \\
		\bottomrule
	\end{tabular}
\end{table*}


In terms of user experience pass rate, our in-house model achieves 83.3\%, significantly higher than the competitor’s 60\%. This indicates that, despite using more tokens and generating more detailed and coherent responses, our model better captures the attribute definitions specified in test cases—such as population group, thermal preference, and health conditions—thereby meeting the UX criteria defined by professional testers more effectively. In contrast, the competitor’s model, constrained by fewer tokens and lower information density, often omits critical attributes, provides vague descriptions, or fails to recognize key health states (e.g., “menstruation” or “rhinitis”), resulting in lower compliance with complex personalization requirements and a reduced pass rate.

Regarding inference latency, our in-house model averages 3.33 seconds, notably slower than the competitor’s 0.51 seconds. This gap primarily stems from the larger number of tokens generated by our model, which requires more extensive semantic reasoning and context processing to ensure comprehensive and attribute-accurate outputs. The competitor’s model prioritizes speed by producing concise, minimal responses—at the cost of completeness and personalization accuracy. Therefore, deployment strategies must balance responsiveness against experience quality: in latency-sensitive scenarios (e.g., real-time voice commands), optimizing inference efficiency for our model may be warranted; whereas in health-conscious or high-personalization contexts (e.g., nighttime sleep mode), the current advantage in user experience pass rate is more valuable.

\subsection{Experiments of Inference and Planning Module}
This section presents experiments conducted separately using a competitor’s closed-source model and our in-house developed model. The evaluation metrics include: (1) the attribute-level consistency rate (i.e., accuracy) between model outputs and the ground-truth labels in the test data, and (2) a subjective evaluation score based on user experience (UX) criteria defined by professional testers.
Notably, this section generates a chain-of-thought (CoT) output, which is derived from a synthesis of the input and output data. Therefore, the consistency rate specifically measures agreement on high-level device control attributes in the final output, while the UX pass rate also evaluates whether the generated chain-of-thought aligns with the experience standards established by testers.
The test dataset consists of data tuples, each comprising environmental conditions, user health status, and current device states. Experimental results are reported under two settings: without and with a post-processing module, enabling a comparative analysis of their respective impacts.


\begin{table*}[htbp]
	\centering
	\caption{Evaluation Criteria with Weighted Scoring (Part A: Rules 1–13)}
	\label{tab:eval_a}
	\begin{tabular}{@{}lclp{6.2cm}@{}}
		\toprule
		\textbf{No.} & \textbf{Dimension} & \textbf{Weight (\%)} & \textbf{Scoring Rules} \\
		\midrule
		1 & Check air params \& user status & 1 &
		1: Consistent; 0: Inconsistent \\
		2 & Temp \& wind speed req. & 5 &
		5: Full match; 3: Partial; 0: Ignored/contradicted \\
		3 & Air quality thresholds & 1 &
		1: Correct threshold; 0: Wrong \\
		4 & Basic AC plan & 5 &
		5: Reasonable; 3: Slight deviation; 0: Contradictory \\
		5 & Auxiliary function plan & 5 &
		5: Correct; 3: Wrong level; 0: Missing \\
		6 & Activation interval & 4 &
		4: 2$\pm$1h, correct; 2: Interval wrong; 0: No interval \\
		7 & \texttt{cmd.mode} & 5 &
		5: Match; 3: Suboptimal; 0: Contradicted \\
		8 & \texttt{cmd.temperature} & 5 &
		5: $\pm$3$^\circ$C of season norm; 3: Outside; 0: Illogical \\
		9 & \texttt{cmd.wind\_speed} & 5 &
		5: Reasonable; 3: Acceptable; 0: Cold-sensitive + high fan \\
		10 & \texttt{cmd.wind\_sensation} & 5 &
		5: Valid; 3: Acceptable; 0: e.g., no-wind in non-cooling \\
		11 & \texttt{cmd.air\_fresh} & 10 &
		10: Correct; 5: Wrong level; 0: Missing \\
		12 & \texttt{cmd.air\_purification} & 10 &
		10: Correct; 5: Wrong level; 0: Missing \\
		13 & \texttt{cmd.air\_humidification} & 10 &
		10: Correct; 5: Wrong level; 0: Missing \\
		\bottomrule
	\end{tabular}
\end{table*}

\begin{table*}[htbp]
	\centering
	\caption{Evaluation Criteria with Weighted Scoring (Part B: Rules 14–25)}
	\label{tab:eval_b}
	\begin{tabular}{@{}lclp{6.2cm}@{}}
		\toprule
		\textbf{No.} & \textbf{Dimension} & \textbf{Weight (\%)} & \textbf{Scoring Rules} \\
		\midrule
		14 & \texttt{cmd.air\_sterilization} & 10 &
		10: On when needed; 8: Unnecessary on; 0: Off when needed \\
		15 & \texttt{threshold.CO2} & 1 &
		1: Correct; 0: Wrong \\
		16 & \texttt{threshold.PM2.5} & — & — \\
		17 & \texttt{threshold.TVOC} & — & — \\
		18 & \texttt{threshold.formaldehyde} & — & — \\
		19 & \texttt{threshold.humidity\_lower} & — & — \\
		20 & \texttt{threshold.humidity\_upper} & — & — \\
		21 & \texttt{interval\_time.air\_fresh} & 2 &
		2: 2$\pm$1h; 1: Interval given but wrong; 0: No interval \\
		22 & \texttt{interval\_time.air\_purification} & 2 &
		2: 2$\pm$1h; 1: Wrong; 0: No interval \\
		23 & \texttt{interval\_time.air\_humidification} & 2 &
		2: 2$\pm$1h; 1: Wrong; 0: No interval \\
		24 & \texttt{interval\_time.air\_sterilization} & 2 &
		2: 2$\pm$1h; 1: Wrong; 0: No interval \\
		25 & \texttt{cmd.tips} (Health Advice) & 10 &
		10: Relevant + caring; 5: Weak/missing care; 0: Missing/wrong \\
		\bottomrule
	\end{tabular}
\end{table*}

%
%
%
%
%
%

This scoring rubric is a fine-grained, multi-dimensional evaluation framework designed specifically for intelligent air conditioning control systems, with the following key characteristics:

(1) User Health and Experience as the Core Focus  
The criteria place strong emphasis on individual user attributes—such as population group, thermal preference, and health conditions—and assign high weights to health-oriented features (e.g., health tips and air treatment functions like sterilization, each weighted at 10\%). This reflects a strategic shift from mere “device control” to “health-conscious service.”

(2) Consistency with Input and Adherence to Common Sense  
The system must not only accurately interpret inputs (e.g., environmental data and user status) but also generate outputs that align with real-world logic and physical constraints—for instance, not setting a temperature in fan-only mode or avoiding high fan speed for users who are very cold-sensitive. Basic functionality (Dimensions 4–10) accounts for 31\% of the total score, ensuring safety and usability.

(3) Differentiation Between “Triggering” and “Correct Execution”  
For auxiliary air treatment functions—fresh air, purification, humidification, and sterilization (Dimensions 11–14)—a three-tier scoring scheme is applied:  
Full points: correctly triggered with appropriate level;  
Partial credit: triggered but with incorrect level;  
Zero points: failed to activate despite meeting trigger conditions.  
This encourages models not just to “recognize” a need, but to “act appropriately.”

(4) Balancing System Compliance and Operational Realism  
Fixed thresholds (e.g., for CO2, PM2.5) must be strictly followed (Dimensions 3, 15–20), ensuring system reliability;  
All auxiliary functions must specify activation intervals (Dimensions 6, 21–24), discouraging unrealistic behaviors like continuous operation for >10 hours, which aligns with typical residential usage patterns.

(5) Thoughtful Weight Allocation Highlighting Critical Capabilities  
Air quality–related functions (fresh air, purification, humidification, sterilization) collectively account for 40\%, underscoring the priority of healthy indoor air;  
Core control parameters (mode, temperature, wind speed) contribute 15\%, guaranteeing fundamental correctness;  
Overall, the weighting balances safety, comfort, intelligence, and practical deployability.

In summary, this rubric evaluates not only the model’s language generation capability but more importantly its decision-making rationality and product readiness in real-world, health-aware smart home scenarios, making it well-suited for assessing AIoT systems with environmental awareness and personalized health care features.

\begin{table*}[htbp]
	\centering
	\caption{Performance Comparison: Competitor’s Model vs. In-house Model}
	\label{tab:performance_comparison}
	\begin{tabular}{lcc}
		\toprule
		\textbf{Metric} & \textbf{Competitor’s Model} & \textbf{In-house Model} \\
		\midrule
		User Experience Pass Rate & 40\% & 94.9\% \\
		Average Inference Latency & 4.82 s & 4.51 s \\
		\bottomrule
	\end{tabular}
\end{table*}

The data shows that the in-house model significantly outperforms the competitor’s model in user experience pass rate (94.9\% vs. 40\%), indicating that its outputs better align with the UX criteria defined by professional testers and more accurately interpret user health status and environmental context to generate appropriate AC control strategies. Moreover, the in-house model achieves a slightly lower average inference latency (4.51s vs. 4.82s), delivering higher accuracy with faster response speed. This demonstrates that the in-house model not only leads substantially in functional correctness but also holds an edge in inference efficiency.

\begin{table*}[htbp]
	\centering
	\caption{Top Error-Prone Evaluation Rules and Scoring Criteria}
	\label{tab:top_rules}
	\begin{tabularx}{\textwidth}{l c L L}
		\toprule
		\textbf{Rule} & \textbf{Weight (\%)} & \textbf{Common Deduction Reason} & \textbf{Scoring Rules} \\
		\midrule
		
		Rule 14: \texttt{cmd.air\_sterilization} (Sterilization) & 23.2\% & 
		Should be ON but was OFF & 
		10 pts: Activated when respiratory illness or regional epidemic is present; otherwise OFF \\
		& & & 
		8 pts: No illness/epidemic but function is ON \\
		& & & 
		0 pts: Illness/epidemic present but function is OFF \\
		
		Rule 4: Basic AC Functional Plan & 20.9\% & 
		\texttt{cmd} inconsistent with or only partially matches ground truth & 
		5 pts: Reasonable and consistent with input \\
		& & & 
		3 pts: Generally consistent (e.g., temperature or fan speed slightly off) \\
		& & & 
		0 pts: Illogical or functionally contradictory (e.g., enables cooling when heating is required; sets temperature in fan-only mode) \\
		
		Rule 7: \texttt{cmd.mode} (Operating Mode) & 12.1\% & 
		Mode generally matches input & 
		5 pts: Fully consistent with input \\
		& & & 
		3 pts: Generally consistent (e.g., heating/dehumidifying condition met but fails to prioritize user’s thermal preference or health state) \\
		& & & 
		0 pts: Contradicts input (e.g., outputs heating when cooling is required) \\
		
		Rule 10: \texttt{cmd.wind\_sensation} & 12.1\% & 
		Generally reasonable & 
		5 pts: Consistent with common sense and current environmental perception \\
		& & & 
		3 pts: Generally reasonable \\
		& & & 
		0 pts: Illogical or functionally incompatible (e.g., “no-wind sensation” enabled in non-cooling mode) \\
		
		\bottomrule
	\end{tabularx}
\end{table*}

The data reveals that the model’s errors in executing air conditioning control strategies are primarily concentrated in the following four rules, which together account for 68.3\% (23.2\% + 20.9\% + 12.1\% + 12.1\%) of all scoring deductions—highlighting the system’s key weaknesses:
Sterilization function (Rule 14, 23.2\%) is the largest source of point loss, mainly due to “should be ON but was OFF”—i.e., failing to activate the sterilization feature when the user has a respiratory illness or during a regional epidemic. This indicates insufficient sensitivity to health risks, likely caused by inaccurate recognition or failure to link disease-related information, leading to critical gaps in health protection.
Basic functional planning (Rule 4, 20.9\%) errors show that the model’s control commands (e.g., temperature, mode, fan speed) often conflict with input conditions or common sense—for example, setting a temperature in fan-only mode or selecting cooling when heating is required. This reflects a need for stronger logical consistency in core decision-making.
Issues with operating mode (Rule 7, 12.1\%) and wind sensation (Rule 10, 12.1\%) mostly manifest as “generally reasonable but not sufficiently precise”—for instance, choosing a standard heating mode despite the user being “very cold-sensitive,” or enabling “no-wind sensation” in non-cooling modes. While not critical failures, these inaccuracies noticeably degrade the refinement of user experience.
In summary, the model currently exhibits significant shortcomings in triggering health-related functions and maintaining logical consistency in basic control decisions, particularly in understanding and responding to health conditions. Future improvements should prioritize enhancing semantic comprehension of health cues and refining the coordination rules among mode, wind sensation, and temperature settings.

\section{Future Work}

Based on the above analysis, future work should focus on the following directions:
Enhance health-related semantic understanding: Improve the model’s accuracy in recognizing user health conditions (e.g., respiratory illnesses, menstruation) to ensure critical functions like sterilization and air purification are reliably activated when needed.
Strengthen logical consistency in control commands: Enforce coherent constraints among core parameters—such as mode, temperature, fan speed, and wind sensation—to eliminate functionally contradictory or common-sense violations.
Optimize the trade-off between inference efficiency and output quality: Further reduce latency while maintaining high user experience pass rates to support responsive real-time interaction.
Implement dynamic memory and state management: Better handle user-reported health status changes (e.g., “I’ve recovered”) to dynamically update or clear personalized control strategies.
Expand test coverage to complex scenarios: Improve robustness and generalization by evaluating performance on edge cases involving vulnerable groups (e.g., children, elderly) and co-occurring health conditions.

\end{document}